\documentclass[conference]{IEEEtran}

\usepackage{array}
\usepackage{cite}
\usepackage{amsmath}
\usepackage{bbm}
\usepackage{algorithmic}
\usepackage{xcolor}
\usepackage{url}
\usepackage{comment}

\makeatletter
\let\MYcaption\@makecaption
\makeatother
\usepackage[font=footnotesize]{subcaption}
\makeatletter
\let\@makecaption\MYcaption
\makeatother

\ifCLASSINFOpdf
    \usepackage[pdftex]{graphicx}
    \graphicspath{{../pdf/}{../jpeg/}}
    \DeclareGraphicsExtensions{.pdf,.jpeg,.png}
\else
    \usepackage[dvips]{graphicx}
    \graphicspath{{../eps/}}
    \DeclareGraphicsExtensions{.eps}
\fi

\usepackage{tikz}
\usepackage{textcomp}
\usepackage{hyperref}
\usepackage{lipsum}

\newcommand\copyrighttext{%
  \footnotesize \textcopyright 2025 IEEE. Personal use of this material is permitted.
  Permission from IEEE must be obtained for all other uses, in any current or future
  media, including reprinting/republishing this material for advertising or promotional
  purposes, creating new collective works, for resale or redistribution to servers or
  lists, or reuse of any copyrighted component of this work in other works.}
\newcommand\copyrightnotice{%
\begin{tikzpicture}[remember picture,overlay]
\node[anchor=south,yshift=10pt] at (current page.south) {\fbox{\parbox{\dimexpr\textwidth-\fboxsep-\fboxrule\relax}{\copyrighttext}}};
\end{tikzpicture}%
}

\begin{document}
\bstctlcite{IEEEexample:BSTcontrol}

\title{Effects of Net Metering Policies on Distributed Energy Resource Valuation and Operation}

\author{\IEEEauthorblockN{Lane D. Smith\textsuperscript{\textdagger} and Daniel S. Kirschen\textsuperscript{\textdaggerdbl}}
\IEEEauthorblockA{\textsuperscript{\textdagger}Department of Energy Science and Engineering, Stanford University, Stanford, CA, USA\\
\textsuperscript{\textdaggerdbl}Department of Electrical and Computer Engineering, University of Washington, Seattle, WA, USA\\
Emails: ldsmith@stanford.edu, kirschen@uw.edu}}

\maketitle

\copyrightnotice

\begin{abstract}
Net energy metering has been a successful policy for increasing solar generation installations and reducing the costs of photovoltaic arrays for consumers. However, increased maturity of solar technologies and concerns over cost shifts created by net energy metering have recently caused the policy to change its incentives. What once favored behind-the-meter solar generation now is focused on compensating flexible operation. This paper explores the impacts that different net energy metering policies have on commercial consumers with various distributed energy resources. We show that the newest iteration of net energy metering is less beneficial for consumers with only solar generation and instead favors those that pair energy storage with solar. Though shiftable flexible demand offers consumers the ability to operate flexibly, the export prices offered by the latest net energy metering policy provide limited value to flexible demand.
\end{abstract}


%
\IEEEpeerreviewmaketitle

\section*{Nomenclature}
\addcontentsline{toc}{section}{Nomenclature}
\begin{IEEEdescription}[\IEEEusemathlabelsep\IEEEsetlabelwidth{$SOC_{min}$}]
    \item[\textit{Sets and Indices}]
    \item[$\mathcal{N}$] Set of demand-charge-specific time-of-use periods, indexed by $n$.
    \item[$\mathcal{S}$] Set of time steps in which the export price is greater than the energy price, indexed by $s$.
    \item[$\mathcal{T}$] Set of time step increments in the user-defined optimization horizon, indexed by $t$.
    \smallskip
    
    \item[\textit{Decision Variables}]
    \item[$d_{dev}^{dn}(t)$] Downward deviation by flexible demand from the consumer's demand profile.
    \item[$d_{dev}^{up}(t)$] Upward deviation by flexible demand from the consumer's demand profile.
    \item[$d_{max}(n)$] Maximum net demand during period $n$.
    \item[$p_{cha}(t)$] Battery's charging power.
    \item[$p_{dis}^{btm}(t)$] Battery's discharging power used for meeting behind-the-meter demand.
    \item[$p_{dis}^{exp}(t)$] Battery's discharging power used for exports.
    \item[$p_{pv}^{btm}(t)$] Power generated by the photovoltaic (PV) array for meeting behind-the-meter demand.
    \item[$p_{pv}^{exp}(t)$] Power generated by the PV array for exports.
    \item[$J(t)$] Battery's state of charge.
    \item[$\zeta_{net}(t)$] Binary variable that is one when net demand equals zero and zero otherwise.
    \smallskip

    \item[\textit{Parameters}]  
    \item[$d(t)$] Consumer's demand.
    \item[$\overline{d}_{dev}^{dn}(t)$] Maximum down deviation by flexible demand.
    \item[$\overline{d}_{dev}^{up}(t)$] Maximum up deviation by flexible demand.
    \item[$J_{init}$] Battery's initial state of charge.
    \item[$\underline{J}$] Battery's minimum state of charge.
    \item[$\overline{J}$] Battery's maximum state of charge.
    \item[$P_{bes}$] Battery's power rating.
    \item[$P_{pv}$] PV array's power rating.
    \item[$\eta_{bes}^{rte}$] Battery's round-trip efficiency.
    \item[$\eta_{inv}$] PV system's inverter efficiency.
    \item[$\pi_{dem}(n)$] Price for maximum demand during period $n$.
    \item[$\pi_{en}(t)$] Price for energy consumption.
    \item[$\pi_{exp}(t)$] Price offered for eligible grid exports.
    \item[$\Delta$] Length of recovery period for flexible demand.
    \smallskip
    
    \item[\textit{Expressions}]
    \item[$CF_{pv}(t)$] PV array's capacity factor profile.
    \item[$d_{net}(t)$] Net consumer demand.
    \item[$\delta(t, n)$] Constant that equals one when $t$ aligns with period $n$ and zero otherwise

\end{IEEEdescription}

\section{Introduction} \label{introduction}
Net energy metering (NEM) has been a largely successful policy aimed at increasing behind-the-meter solar generation and helping reduce the costs of solar photovoltaic (PV) systems for consumers. However, as solar PVs have become more mature and have decreased in cost, public utilities commissions have begun to update their NEM programs. The original iteration, ``NEM 1.0,'' compensated consumers for any excess generation exports at the same amount as their electricity rate. As concerns over cost shifts arose, ``NEM 2.0'' was introduced, resulting non-bypassable charges, on the order of \$0.02/kWh to \$0.03/kWh, that cannot be offset by NEM credits \cite{cpuc_nem2_release}. With further concerns over cost shifts \cite{borenstein_nature}, the California Public Utilities Commission (CPUC) recently adopted the Net-Billing Tariff (NBT), commonly referred to as ``NEM 3.0,'' which aims to better align export prices with the value of avoided system costs that can be provided \cite{cpuc_net_billing_tariff}.

While NEM policies were originally established to help integrate distributed solar, NEM 3.0 proposals like the CPUC's NBT indicate a shift in the technology of interest. Though a renewable energy generating facility (e.g., solar PVs), is still necessary in NEM programs, the focus on export values less aligned with solar-generating times suggests a push to enable more distributed flexible resources \cite{cpuc_net_billing_tariff}. Solar-plus-storage systems are not novel, but previous NEM policies did little more than encourage storage to partake in arbitrage and peak shaving. New export prices considering system-level value could make exports an additional value stream for properly configured battery energy storage (BES). However, BES is not the only distributed flexible resource. Sources of shiftable flexible demand, such as electric vehicles and heat pumps, can respond to prices too. As such, we must understand how new NEM policies not only impact solar PVs, but also distributed flexible resources like BES and flexible demand.

In this paper, we assess the effects of NEM policies on different consumers with various distributed energy resources (DERs). To adequately explore these impacts, a mathematical optimization model is created that determines the minimum electricity bill for a consumer with a combination of PV, BES, and flexible demand assets. This model allows different combinations of assets, consumer load shapes, and NEM policies to be explored in the PG\&E service territory.

The rest of this paper is organized as follows. Section \ref{mathematical_formulation} describes the mathematical model used to minimize a commercial consumer's electricity bill. Section \ref{case_study} introduces the case studies that examine the impacts of NEM variations on consumers with different behind-the-meter asset mixes. Section \ref{results_and_discussion} presents and discusses the results obtained from the simulations. Section \ref{conclusions} concludes the paper.

\section{Mathematical Formulation} \label{mathematical_formulation}
We define a mixed-integer linear program (MILP) to determine the minimum electricity bill, comprised of costs associated with time-of-use (TOU) rates and revenues associated with NEM, for a consumer with a combination of PV, BES, and shiftable flexible demand resources. Within the MILP, asset operation is optimized to achieve the minimum bill, while a simulated demand profile and a PV capacity factor profile are provided as parameters. The MILP is formulated as follows:
\begin{equation}
    \begin{aligned}
        \min. \quad &\sum_{n \in \mathcal{N}} \pi_{dem} (n) \cdot d_{max}(n) + \sum_{t \in \mathcal{T}} \pi_{en}(t) \cdot d_{net}(t) \\
        &- \sum_{t \in \mathcal{T}} \pi_{exp}(t) \cdot \left[ p_{pv}^{exp}(t) + p_{dis}^{exp}(t) \right]
    \end{aligned}
    \label{eq:ch3_obj}
\end{equation}

\noindent subject to:
\begin{equation}
    d_{net}(t) \geq 0, \; \forall t \in \mathcal{T}
    \label{eq:ch3_net_dem_lb}
\end{equation}
\begin{equation}
    \zeta_{net}(s) = 1 \Rightarrow d_{net}(s) \leq 0, \; \forall s \in \mathcal{S}
    \label{eq:ch3_net_dem_exp_ind_var}
\end{equation}
\begin{equation}
    p_{pv}^{exp}(s) + p_{dis}^{exp}(s) \leq \zeta_{net}(s) \cdot (P_{pv} + P_{bes}), \; \forall s \in \mathcal{S} \subseteq \mathcal{T}
    \label{eq:ch3_net_dem_exp_link}
\end{equation}
\begin{equation}
    \delta (t, n) \cdot d_{net}(t) \leq d_{max}(n), \; \forall t \in \mathcal{T}, \; \forall n \in \mathcal{N}
    \label{eq:ch3_max_dem}
\end{equation}
\begin{equation}
    d_{max}(n) \geq 0, \; \forall n \in \mathcal{N}
    \label{eq:ch3_max_dem_nonneg}
\end{equation}
\begin{equation}
    p_{pv}^{btm}(t) \geq 0, \; \forall t \in \mathcal{T}
    \label{eq:ch3_pv_btm_lb}
\end{equation}
\begin{equation}
    p_{pv}^{exp}(t) \geq 0, \; \forall t \in \mathcal{T}
    \label{eq:ch3_pv_exp_lb}
\end{equation}
\begin{equation}
    p_{pv}^{btm}(t) + p_{pv}^{exp}(t) \leq \eta_{inv} \cdot P_{pv} \cdot CF_{pv}(t), \; \forall t \in \mathcal{T}
    \label{eq:ch3_pv_ub}
\end{equation}
\begin{equation}
    0 \leq p_{cha}(t) \leq P_{bes}, \; \forall t \in \mathcal{T}
    \label{eq:ch3_cha_bounds}
\end{equation}
\begin{equation}
    p_{dis}^{btm}(t) \geq 0, \; \forall t \in \mathcal{T}
    \label{eq:ch3_dis_bound_btm_lb}
\end{equation}
\begin{equation}
    p_{dis}^{exp}(t) \geq 0, \; \forall t \in \mathcal{T}
    \label{eq:ch3_dis_bound_exp_lb}
\end{equation}
\begin{equation}
    p_{dis}^{btm}(t) + p_{dis}^{exp}(t) \leq P_{bes}, \; \forall t \in \mathcal{T}
    \label{eq:ch3_dis_bound_ub}
\end{equation}
\begin{equation}
    J(t) = J(t - 1) + \eta_{bes}^{rte} \cdot p_{cha}(t) - p_{dis}^{btm}(t) - p_{dis}^{exp}(t), \; \forall t \in \mathcal{T} >0
    \label{eq:ch3_soc}
\end{equation}
\begin{equation}
    J(t) = J_{init} + \eta_{bes}^{rte} \cdot p_{cha}(t) - p_{dis}^{btm}(t) - p_{dis}^{exp}(t), \; t=0
    \label{eq:ch3_soc0}
\end{equation}
\begin{equation}
    \underline{J} \leq J(t) \leq \overline{J}, \; \forall t \in \mathcal{T}
    \label{eq:ch3_soc_bounds}
\end{equation}
\begin{equation}
    J(| \mathcal{T} |) \geq J_{init}
    \label{eq:ch3_soc_last}
\end{equation}
\begin{equation}
    p_{cha}(t) \leq p_{pv}^{btm}(t), \; \forall t \in \mathcal{T}
    \label{eq:ch3_no_import}
\end{equation}
\begin{equation}
    0 \leq d_{dev}^{dn}(t) \leq \overline{d}_{dev}^{dn}(t), \; \forall t \in \mathcal{T}
    \label{eq:ch3_flex_dem_down_bounds}
\end{equation}
\begin{equation}
    0 \leq d_{dev}^{up}(t) \leq \overline{d}_{dev}^{up}(t), \; \forall t \in \mathcal{T}
    \label{eq:ch3_flex_dem_up_bounds}
\end{equation}
\begin{equation}
    \sum_{t \in \mathcal{T}} \left[ d_{dev}^{up}(t) - d_{dev}^{dn}(t) \right] = 0
    \label{eq:ch3_interval_dem_bal}
\end{equation}
\begin{equation}
    \sum_{\tau = k}^{\Delta + k - 1} \left[ d_{dev}^{up}(\tau) - d_{dev}^{dn}(\tau) \right] \geq 0, \; k = 1, ..., |\mathcal{T}| - \Delta + 1
    \label{eq:ch3_rolling_dem_bal}
\end{equation}
\noindent where
\begin{equation}
    \begin{aligned}
        d_{net}(t) &= d(t) - p_{pv}^{btm}(t) + p_{cha}(t) - p_{dis}^{btm}(t) \\
        &+ d_{dev}^{up}(t) - d_{dev}^{dn}(t), \; \forall t \in \mathcal{T}
    \end{aligned}
    \label{eq:ch3_net_dem}
\end{equation}

Equation \eqref{eq:ch3_obj} is the objective function, reflecting the consumer's total electricity bill. The first summation pertains to the tariff's demand charges, the second summation describes the tariff's energy charge, and the third summation represents the NEM revenue, where the value of the export price varies depending on the NEM version being considered. Note that BES exports $p_{dis}^{exp}$ may be prohibited depending on the scenario under consideration. Constraint \eqref{eq:ch3_net_dem_lb} prevents the consumer's net demand, which is defined in Equation \eqref{eq:ch3_net_dem} and does not include exports $p_{pv}^{exp}$ and $p_{dis}^{exp}$, from being exported to the grid. Constraints \eqref{eq:ch3_net_dem_exp_ind_var} and  \eqref{eq:ch3_net_dem_exp_link} ensure there are only exports if the consumer's assets have already met the net demand. Differing from \cite{smith_pesgm21} and \cite{smith_pesgm22}, exports and behind-the-meter consumption are disaggregated to prevent this practical infeasibility, which can occur when export prices are greater than energy prices as seen under the CPUC's NBT. Constraints \eqref{eq:ch3_max_dem} and \eqref{eq:ch3_max_dem_nonneg} define the maximum demand during different periods.

Constraints \eqref{eq:ch3_pv_btm_lb} -- \eqref{eq:ch3_pv_ub} describe the PV model's bounds. The PV capacity factor profile is determined by dividing the simulated PV array's generation output by its rated capacity. The PV array's generation output is determined using solution methods for solving the single-diode PV model, with the detailed formulation available in the DERIVE model \cite{derive_repo, smith_dissertation}.

Constraints \eqref{eq:ch3_cha_bounds} -- \eqref{eq:ch3_no_import} describe the BES model. Constraint \eqref{eq:ch3_cha_bounds} restricts the charging power of the BES, while Constraints \eqref{eq:ch3_dis_bound_btm_lb} -- \eqref{eq:ch3_dis_bound_ub} restrict the discharging power. Constraints \eqref{eq:ch3_soc} and \eqref{eq:ch3_soc0} define the state of charge at time $t$. Constraint \eqref{eq:ch3_soc_bounds} enforces upper and lower bounds on the state of charge. Constraint \eqref{eq:ch3_soc_last} ensures that the final state of charge (at $t = | \mathcal{T} |$) is no less than the initial state of charge, preventing perverse discharging schemes near the later time steps of the optimization horizon. Constraint \eqref{eq:ch3_no_import} prevents the BES from charging from the grid, if specified by the scenario.

Constraints \eqref{eq:ch3_flex_dem_down_bounds} -- \eqref{eq:ch3_rolling_dem_bal} describe the shiftable flexible demand model. Constraints \eqref{eq:ch3_flex_dem_down_bounds} and \eqref{eq:ch3_flex_dem_up_bounds} bound the amount demand deviates from the consumer's baseline demand. Constraint \eqref{eq:ch3_interval_dem_bal} ensures demand is balanced over the optimization horizon. Constraint \eqref{eq:ch3_rolling_dem_bal} is a rolling window of size $\Delta$ in which demand deviations cannot result in a net decrease in demand.

\section{Case Study Design} \label{case_study}
We consider two archetypal commercial consumers to assess the impacts of different NEM policies: a consumer with morning-and-evening-peaking (MEP) demand and a consumer with midday-peaking (MDP) demand. These consumers are modeled using the Department of Energy's large hotel and supermarket commercial prototype models, respectively, and typical meteorological year (TMY3) data from Fresno Yosemite International Airport in Fresno, CA. The demand profiles are available from \cite{openei} and have maximum demands of 444 kW and 358 kW for the MEP and MDP consumers.

With a source of renewable energy generation required for participation under NEM programs, each scenario models the consumers alongside a simulated solar PV system \cite{nem2}. When testing the impact of different NEM policies on solar-only consumers, the PV's rated capacity is set to values between 0\% and 150\% of the consumers' maximum demand. When included alongside BES or flexible demand, the PV's rated capacity is sized according to the consumers' maximum demand. To determine the PV's capacity factor profile, we use TMY3 data from Fresno Yosemite International Airport \cite{comstock}.

To explore the impact of different NEM policies on consumers with flexible resources, we consider two configurations: PVs paired with either BES or flexible demand. The BES's rated power capacity is set to values between 0\% and 150\% of the consumers' maximum demand, with durations of two and six hours being considered. We model three management schemes for solar-plus-storage consumers. The first allows BES to charge from the grid, but prohibits grid exports. The second only allows BES to charge from paired PVs, but allows exports to the grid. Both designs are commonly allowed in NEM programs \cite{nem2, zinaman_nrel}. The third, which is atypical for NEM, allows BES to charge from and export to the grid.

For flexible demand, we consider demand flexibility percentages ranging from 0\% to 100\% and load recovery periods of two, six, twelve, and twenty-four hours. We define ``demand flexibility percentage'' as the percent of the base demand that can be curtailed or increased during a given time step, and we use it to populate the bounds placed on the demand deviations. As noted in \cite{smith_pesgm22}, potential bottlenecks during low-demand periods can cause this model to be conservative, but we ultimately believe it is a useful consideration for modeling technology-agnostic flexible demand that may experience intermittent availability from its portfolio of assorted assets.

The modeled consumers are exposed to PG\&E's Electric Schedule B-19 rate, a TOU tariff offered to commercial consumers \cite{b19}. We consider four NEM policies: NEM 1.0, NEM 2.0, NEM 3.0, and no NEM. NEM 1.0 sets export prices equal to the TOU energy prices. NEM 2.0 sets the effective export price signal equal to the TOU energy prices minus a non-bypassable charge of \$0.02977/kWh \cite{nem2}. NEM 3.0 mimics the CPUC's NBT, which stipulates that export prices are to equal the outputs from E3's Avoided Cost Calculator \cite{e3_acc}, averaged by month, weekdays versus weekends and holidays, and hour of the day \cite{cpuc_net_billing_tariff}. Scenarios without NEM prohibit exports, causing PV generation to be consumed or curtailed.

\section{Results and Discussion} \label{results_and_discussion}
The model from Section \ref{mathematical_formulation} is implemented in DERIVE \cite{derive_repo}, an open-source DER valuation tool we use to run the simulations. The following subsections present results from these simulations and discuss how different NEM policies affect asset valuation.

\subsection{Impacts on Consumers with Only Solar PVs} \label{solar_only_results}

\begin{figure}
    \centering
    \begin{subfigure}[b]{0.24\textwidth}
         \centering
         \includegraphics[width=\textwidth]{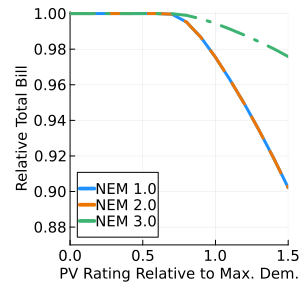}
         \caption{}
        \label{subfig:bill_mep_solar_only}
     \end{subfigure}
     \hfill
     \begin{subfigure}[b]{0.24\textwidth}
         \centering
         \includegraphics[width=\textwidth]{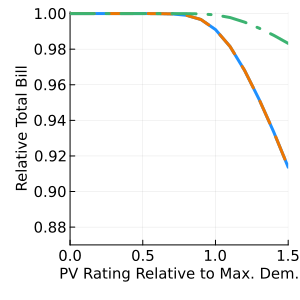}
         \caption{}
        \label{subfig:bill_mdp_solar_only}
     \end{subfigure}
    \caption{Total electricity bill under different NEM policies relative to that under no NEM for different consumers with only PVs. (a) depicts the MEP consumer's relative bills and (b) depicts the same for the MDP consumer.}
    \label{fig:pv_only_value}
\end{figure}

To understand the impact of different NEM policies on PV systems, we first examine participation by solar-only consumers with different-sized systems. Numerical results can be seen in Figure \ref{fig:pv_only_value}, which shows the total electricity bills under NEM 1.0, 2.0, and 3.0 relative to the total electricity bill under no NEM participation for solar-only consumers.

For PV systems that are not large enough to produce excess generation, all generation is self-consumed. From Figure \ref{fig:pv_only_value}, we can see that the MEP consumer requires a smaller relative PV capacity compared to the MDP consumer before excess generation can be exported. This is due in large part to the consumers' load shapes, with the MEP consumer having a smaller relative demand that coincides with PV generation compared to that of the MDP consumer.

Once the PV systems are relatively large enough to produce exports, it is clear that NEM 1.0 and 2.0 provide consumers with the greatest export compensation while NEM 3.0 provides the worst; had larger PV capacities been considered, situations would arise where a difference between NEM 1.0 and 2.0 compensation could be observed. Though NEM 3.0 offers export prices that can vastly exceed those offered under NEM 1.0 (a maximum export prices of \$2.96644/kWh and \$0.21585/kWh, respectively), such prices typically occur in the late afternoon and early evening when PV generation wanes. With no flexible resources to take advantage of these export prices, NEM 1.0 can offer consumers better value during solar-generating hours, as NEM 1.0 provides a greater export price than NEM 3.0 nearly 95\% of the time during those hours.

\subsection{The Value of Battery Energy Storage Systems} \label{solar_plus_storage_results}

\begin{figure*}
     \centering
     \begin{subfigure}[b]{0.24\textwidth}
         \centering
         \includegraphics[width=\textwidth]{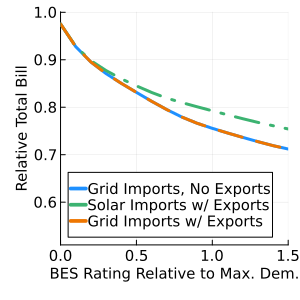}
         \caption{}
        \label{subfig:MEP_2h_BES_NEM_2}
     \end{subfigure}
     \hfill
     \begin{subfigure}[b]{0.24\textwidth}
         \centering
         \includegraphics[width=\textwidth]{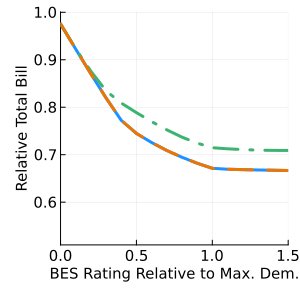}
         \caption{}
        \label{subfig:MEP_6h_BES_NEM_2}
     \end{subfigure}
     \hfill
     \begin{subfigure}[b]{0.24\textwidth}
         \centering
         \includegraphics[width=\textwidth]{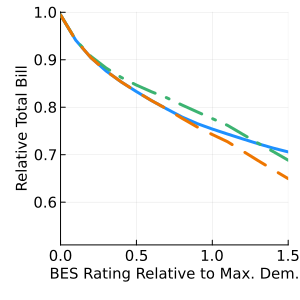}
         \caption{}
        \label{subfig:MEP_2h_BES_NEM_3}
     \end{subfigure}
     \hfill
     \begin{subfigure}[b]{0.24\textwidth}
         \centering
         \includegraphics[width=\textwidth]{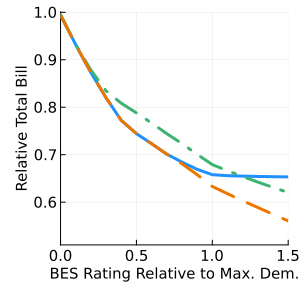}
         \caption{}
        \label{subfig:MEP_6h_BES_NEM_3}
     \end{subfigure}
     \\
     \begin{subfigure}[b]{0.24\textwidth}
         \centering
         \includegraphics[width=\textwidth]{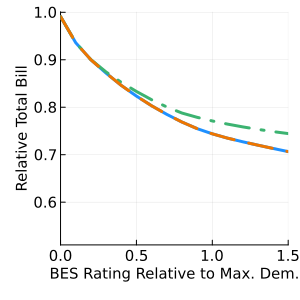}
         \caption{}
        \label{subfig:MDP_2h_BES_NEM_2}
     \end{subfigure}
     \hfill
     \begin{subfigure}[b]{0.24\textwidth}
         \centering
         \includegraphics[width=\textwidth]{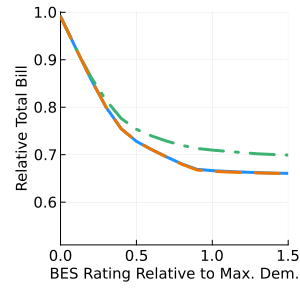}
         \caption{}
        \label{subfig:MDP_6h_BES_NEM_2}
     \end{subfigure}
     \hfill
     \begin{subfigure}[b]{0.24\textwidth}
         \centering
         \includegraphics[width=\textwidth]{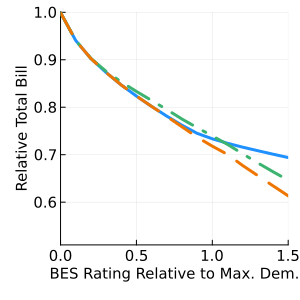}
         \caption{}
        \label{subfig:MDP_2h_BES_NEM_3}
     \end{subfigure}
     \hfill
     \begin{subfigure}[b]{0.24\textwidth}
         \centering
         \includegraphics[width=\textwidth]{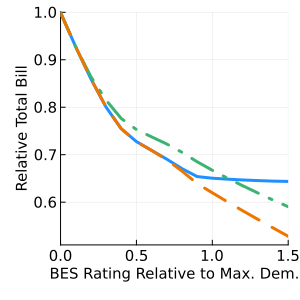}
         \caption{}
        \label{subfig:MDP_6h_BES_NEM_3}
     \end{subfigure}
        \caption{Total electricity bill under different NEM policies and BES management schemes for consumers with PVs and BES relative to that for consumers under no NEM with only PVs. (a) -- (d) depict the MEP consumer's relative bills and (e) -- (h) depict the same for the MDP consumer. (a), (b), (e), and (f) show consumer participation under NEM 2.0 and (c), (d), (g), and (h) show participation under NEM 3.0. (a), (c), (e), and (g) depict a two-hour BES and (b), (d), (f), and (h) depict a six-hour BES. Each figure shows a range of BES power ratings (relative to the consumers' maximum demand).}
        \label{fig:bes_value}
\end{figure*}

As was discussed in the previous subsection, the inclusion of distributed flexibility can better help consumers take advantage of export prices offered under NEM policies, especially those offered by NEM 3.0. To better understand the value of distributed flexibility, we first look at BES of different capacities and durations paired with PVs rated to meet the consumers' maximum demand. Results from this analysis are shown in Figure \ref{fig:bes_value}, which shows the total electricity bill for consumers with BES and PVs under different NEM policies and different BES management schemes relative to the total electricity bill under no NEM participation for a solar-only consumer.

From Figures \ref{subfig:MEP_2h_BES_NEM_2}, \ref{subfig:MEP_6h_BES_NEM_2}, \ref{subfig:MDP_2h_BES_NEM_2}, and \ref{subfig:MDP_6h_BES_NEM_2}, we can see the value of adding BES to a PV system when participating under NEM 2.0. Under a NEM policy with effective export prices equal to or less than the TOU energy price, there is no value in exporting energy to the grid. With the flexibility of BES, energy that would otherwise be exported can be shifted to a later, more-expensive TOU period. As such, despite the ability of BES to both charge from and export to the grid in one of the modeled scenarios, we observe an identical relative total bill in the scenario where BES can charge from but cannot export to the grid. Both of these scenarios also result in lower relative total bills than the scenario in which the BES can only charge from its PVs, but is allowed to export to the grid. This latter scenario loses out on valuable arbitrage opportunities afforded to consumers under the other two BES management schemes. However, as can be seen in the figures, the value of the BES plateaus as its power rating increases, indicating that there is limited arbitrage and peak-shaving value available.

Under NEM 3.0, which influences the responses shown in Figures \ref{subfig:MEP_2h_BES_NEM_3}, \ref{subfig:MEP_6h_BES_NEM_3}, \ref{subfig:MDP_2h_BES_NEM_3}, and \ref{subfig:MDP_6h_BES_NEM_3}, high export prices provide an additional value stream. For smaller relative BES ratings, it is most valuable to participate in arbitrage and peak shaving, as evidenced by the identical relative total bills obtained using the two grid-imports-oriented management schemes. However, as the relative BES rating increases, the ability for the BES to export to the grid becomes increasingly valuable, especially as the value from arbitrage and peak shaving becomes exhausted. While charging from the grid and being able to export allows for the lowest relative total bill, the two more-realistic management schemes each can provide consumers valuable bill savings depending on the BES rating relative to the PV rating.

\subsection{Limited Value for Non-Storage Flexible Resources} \label{solar_plus_demand_flexibility_results}

\begin{figure*}
     \centering
     \begin{subfigure}[b]{0.24\textwidth}
         \centering
         \includegraphics[width=\textwidth]{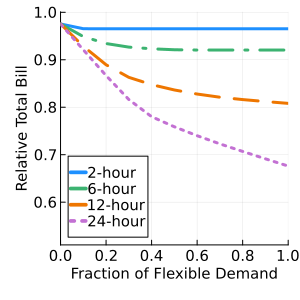}
         \caption{}
        \label{subfig:MEP_flex_dem_NEM_2}
     \end{subfigure}
     \hfill
     \begin{subfigure}[b]{0.24\textwidth}
         \centering
         \includegraphics[width=\textwidth]{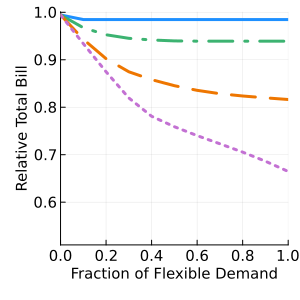}
         \caption{}
        \label{subfig:MEP_flex_dem_NEM_3}
     \end{subfigure}
     \hfill
     \begin{subfigure}[b]{0.24\textwidth}
         \centering
         \includegraphics[width=\textwidth]{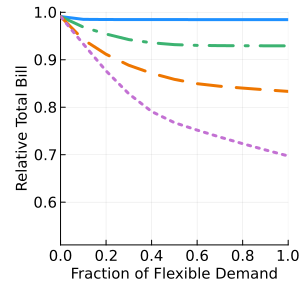}
         \caption{}
        \label{subfig:MDP_flex_dem_NEM_2}
     \end{subfigure}
     \hfill
     \begin{subfigure}[b]{0.24\textwidth}
         \centering
         \includegraphics[width=\textwidth]{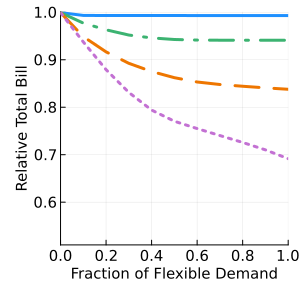}
         \caption{}
        \label{subfig:MDP_flex_dem_NEM_3}
     \end{subfigure}
        \caption{Total electricity bill under different NEM policies for consumers with PVs and flexible demand relative to that for consumers under no NEM with only PVs. (a) and (b) depict the MEP consumer's relative bills and (c) and (d) depict the same for the MDP consumer. (a) and (c) show consumer participation under NEM 2.0 and (b) and (d) show participation under NEM 3.0. Each figure shows various recovery periods and a range of demand flexibility percentages.}
        \label{fig:flex_dem_value}
\end{figure*}

Though flexible demand is less suited for NEM programs compared to BES, the ability to shift demand can enable the PV system to export during times with high export prices. We examine pairing PVs and flexible demand with different demand flexibility percentages and recovery durations to understand the value provided by another source of flexibility. Figure \ref{fig:flex_dem_value} shows the total electricity bill for consumers with flexible demand and PVs under different NEM policies relative to that under no NEM participation for a solar-only consumer.

From Figure \ref{fig:flex_dem_value}, we can see that the relative total bills for each consumer are similar regardless of participation under NEM 2.0 or NEM 3.0. This indicates that most of the value is not derived from the export prices, but rather from price differentials in the underlying TOU rate. Only at its most flexible (i.e., a demand flexibility percentage near 100\% and a recovery duration near twenty-four hours) does it appear that flexible demand is able to benefit under NEM 3.0. This is in stark contrast to BES, which saw benefits under each management scheme with the shift from NEM 2.0 to 3.0.

\section{Conclusions} \label{conclusions}
This paper shows the impact that different NEM policies have on consumers with different load shapes and different DER mixes. The transition from NEM 2.0 to NEM 3.0, modeled after CPUC's NBT, will negatively impact consumers with only solar PVs. The inclusion of large, but limited, export prices in NEM 3.0 necessitate that consumers have a specific source of distributed flexibility at their disposal. BES, under any of the three explored management schemes, can provide increased value to consumers under NEM 3.0 compared to NEM 2.0. However, consumers with flexible demand are unable to see the same value gains between NEM 2.0 and 3.0, even if flexible demand is at its most flexible. These results highlight another example of the CPUC and PG\&E prioritizing flexibility from BES over that from flexible demand. Rather than offering technology-specific rate structures \cite{smith_pesgm22} or discriminatory programs, such as CPUC's NBT, it would behoove regulators to offer price signals that can appropriately tap the value of flexible demand. Previous proposals for implementing symmetrical rates \cite{perez-arriaga_framework} or rates that better enable real-time price discovery alongside transactive principles \cite{pnnl_dsot} could help make flexible demand more competitive with other DERs.

Future work will focus on designing rate structures that better extract flexibility from technology-agnostic resources. Additionally, future work will explore computationally efficient models for exporting excess energy when the export price is greater than the TOU energy price.

\bibliographystyle{IEEEtran}
\bibliography{main.bbl}

\begin{thebibliography}{10}
\providecommand{\url}[1]{#1}
\csname url@samestyle\endcsname
\providecommand{\newblock}{\relax}
\providecommand{\bibinfo}[2]{#2}
\providecommand{\BIBentrySTDinterwordspacing}{\spaceskip=0pt\relax}
\providecommand{\BIBentryALTinterwordstretchfactor}{4}
\providecommand{\BIBentryALTinterwordspacing}{\spaceskip=\fontdimen2\font plus
\BIBentryALTinterwordstretchfactor\fontdimen3\font minus
  \fontdimen4\font\relax}
\providecommand{\BIBforeignlanguage}[2]{{%
\expandafter\ifx\csname l@#1\endcsname\relax
\typeout{** WARNING: IEEEtran.bst: No hyphenation pattern has been}%
\typeout{** loaded for the language `#1'. Using the pattern for}%
\typeout{** the default language instead.}%
\else
\language=\csname l@#1\endcsname
\fi
#2}}
\providecommand{\BIBdecl}{\relax}
\BIBdecl

\bibitem{cpuc_nem2_release}
{California Public Utilities Commission}, ``{CPUC Issues Proposed Net Energy
  Metering Decision to Ensure Customers Continue to Benefit from Going
  Solar},'' 2015, {Press Release. Docket \#: R.14-07-002}.

\bibitem{borenstein_nature}
S.~Borenstein, ``{It's time for rooftop solar to compete with other
  renewables},'' \emph{Nature Energy}, vol.~7, p. 298, 2022.

\bibitem{cpuc_net_billing_tariff}
{California Public Utilities Commission}, ``{Decision Revising Net Energy
  Metering Tariff and Subtariffs},'' 2022, {Decision 22-12-056}.

\bibitem{smith_pesgm21}
L.~D. Smith and D.~S. Kirschen, ``{Impacts of Time-of-Use Rate Changes on the
  Electricity Bills of Commercial Consumers},'' in \emph{IEEE Power \& Energy
  Society General Meeting}, 2021.

\bibitem{smith_pesgm22}
L.~D. Smith and D.~S. Kirschen, ``{Should Storage-Centric Tariffs be Extended
  to Commercial Flexible Demand?}'' in \emph{IEEE Power \& Energy Society
  General Meeting}, 2022.

\bibitem{derive_repo}
\BIBentryALTinterwordspacing
L.~D. Smith. {``DERIVE''}. GitHub Repository. [Online]. Available:
  \url{https://github.com/lanesmith/DERIVE}
\BIBentrySTDinterwordspacing

\bibitem{smith_dissertation}
L.~D. Smith, ``{Electricity Rate Design for Integrating Distributed Resources
  Into Energy Systems},'' Ph.D. dissertation, University of Washington, 2024.

\bibitem{openei}
\BIBentryALTinterwordspacing
S.~Ong and N.~Clark. (2022) {``Commercial and Residential Hourly Load Profiles
  for all TMY3 Locations in the United States''}. OpenEI. [Online]. Available:
  \url{https://data.openei.org/submissions/153}
\BIBentrySTDinterwordspacing

\bibitem{nem2}
\BIBentryALTinterwordspacing
{Pacific Gas and Electric Company}. (2023) {Electric Schedule NEM2: Net Energy
  Metering Service}. Accessed on: 2023-11-16. [Online]. Available:
  \url{https://www.pge.com/tariffs/assets/pdf/tariffbook/ELEC_SCHEDS_NEM2.pdf}
\BIBentrySTDinterwordspacing

\bibitem{comstock}
A.~Parker, H.~Horsey, M.~Dahlhausen, M.~Praprost, C.~CaraDonna, A.~LeBar, and
  L.~Klun, ``{ComStock Reference Documentation: Version 1},'' National
  Renewable Energy Laboratory, Golden, CO, Tech. Rep. NREL/TP-5500-83819, 2023.

\bibitem{zinaman_nrel}
O.~Zinaman, T.~Bowen, and A.~Aznar, ``{An Overview of Behind-the-Meter
  Solar-Plus-Storage Regulatory Design},'' National Renewable Energy
  Laboratory, Golden, CO, Tech. Rep., 2020.

\bibitem{b19}
\BIBentryALTinterwordspacing
{Pacific Gas and Electric Company}. (2023) {Electric Schedule B-19: Medium
  General Demand-Metered TOU Service}. Accessed on: 2023-11-14. [Online].
  Available:
  \url{https://www.pge.com/tariffs/assets/pdf/tariffbook/ELEC_SCHEDS_B-19.pdf}
\BIBentrySTDinterwordspacing

\bibitem{e3_acc}
{Energy and Environmental Economics, Inc.}, ``{2022 Distributed Energy
  Resources Avoided Cost Caluclator Documentation},'' 2022.

\bibitem{perez-arriaga_framework}
I.~J. P\'erez-Arriaga, J.~D. Jenkins, and C.~Batlle, ``{A regulatory framework
  for an evolving electricity sector: Highlights of the MIT utility of the
  future study},'' \emph{Economics of Energy \& Environmental Policy}, vol.~6,
  no.~1, pp. 71--92, 2017.

\bibitem{pnnl_dsot}
H.~M. Reeve, S.~Widergren, R.~Pratt, B.~Bhattarai, S.~Hanif, S.~Bender,
  T.~Hardy, and M.~Pelton, ``{Distribution System Operator with Transactive
  (DSO+T) Study: Main Report},'' Pacific Northwest National Laboratory,
  Richland, WA, Tech. Rep. PNNL-32170-1, 2022.

\end{thebibliography}

\end{document}